\documentclass{elsart}
\usepackage{bm,amsmath,amssymb,latexsym,mathrsfs,enumerate,graphicx,subfig}
\usepackage[numbers,sort&compress]{natbib}
\usepackage{hyperref}
\renewcommand{\vec}[1]{\bm{#1}}
\begin{document}
\begin{frontmatter}
\title{Equilibrium magnetisation structures in ferromagnetic nanorings}
\author{Volodymyr P. Kravchuk},
\author{Denis D. Sheka\corauthref{cor}},
\corauth[cor]{Corresponding author.} \ead{denis\_sheka@univ.kiev.ua}
\address{National Taras Shevchenko University of Kiev, 03127 Kiev, Ukraine}
\author{Yuri B. Gaididei}
\address{Institute for Theoretical Physics, 03143 Kiev, Ukraine}

\date{April 8, 2006}

\begin{abstract}
The ground state of the ring--shape magnetic nanoparticle is
studied. Depending on the geometrical and magnetic parameters of the
nanoring, there exist different magnetisation configurations
(magnetic phases): two phases with homogeneous magnetisation
(easy--axis and easy--plane phases) and two inhomogeneous (planar
vortex phase and out--of--plane one). The existence of a new
intermediate out--of--plane vortex phase, where the inner
magnetisation is not strongly parallel to the easy axis, is
predicted. Possible transitions between different phases are
analysed using the combination of analytical calculations and
micromagnetic simulations.
\end{abstract}
\end{frontmatter}

\section{Introduction}
\label{sec:intro}

Magnetic nanoparticles and their structures have become a subject of
interest in the past few years \cite{Skomski03}. For nanoparticles
of axial geometry, such as disks (\emph{nanodots}) and rings
(\emph{nanorings}), the vortex configuration becomes the lowest
energy state when the particle diameter exceeds the single--domain
size owing to the competition between exchange and magnetostatic
interactions \cite{Hubert98}. Because of nontrivial topological
properties, the vortex state nanoparticles are promising candidates
for the high--density magnetic storage devices, some designs of
magnetic memories and high--resolution magnetic field sensors
\cite{Cowburn02}.

There are known two types of vortices in a nanomagnetism. For the
disk--shaped particle there exist an out--of--plane vortex (OPV),
which is characterised by the core magnetisation perpendicular to
the disk plane. Different static and dynamical properties of
magnetic nanodisks have been studied recently. In particular,
theoretical and experimental studies of the transition between
single domain and vortex configurations (phases)
\cite{Cowburn99,Cowburn02,Hoellinger03,Scholz03,Ross02} give a
possibility to conclude, that the vortex state in a nanodisk can
exists for the dot diameter in the submicron range, which is much
more than the minimum nanoparticle size, currently used in
experiments. That makes serious limitations for considering such
structure as competitive one for the magnetisation storage.

The ring shape potentiates more magnetisation configurations, see
for review \cite{Klaui03a}. It is well--known \cite{Klaui03a} that
the pure in--plane vortex (IPV) configuration can be realised in the
nanoring, where the perpendicular magnetisation component is absent.
Such a vortex has lower energy than the OPV in a disc, it does not
produce surface magnetostatic charges, thus it can provides more
stable magnetic bit for the magnetisation storage.

Since there exist pure OPV for the disk and pure IPV for the ring,
there appears a reasonable question, whether the intermediate state
exists or not. We will show in the paper that \emph{intermediate
vortices} (IMV) exist in the nanoring when the inner hole is small
enough. For such vortices the magnetisation inside the vortex core
has an out--of--plane component, but even in the vortex centre it is
not perpendicular to the ring plane. The presence of the hole in the
disk decreases the ``singularity'' in the vortex centre and so it
gives an opportunity to realise the vortex state in a smaller
nanoparticle. We will demonstrate that the vortex state can be
realised in a nanoring with diameter of few tens of nanometers.

In this paper we provide the systematic study of the magnetic
nanoring's ground state. Namely, by combining analytical methods and
micromagnetic simulations we describe the equilibrium magnetic
phases: two homogeneous states with magnetisation directions along
the ring plane and perpendicular to it, and inhomogeneous vortex
states, including OPV, IPV, and IMV states. The key moment is to
propose the simple analytical approach, which allows us to describe
different states and transitions between them.

The paper is organised as follows. In Sec.~\ref{sec:model} we
formulate the model and treat it analytically for the homogeneously
magnetised nanoring. The vortex--state nanorings are analysed in
Sec.~\ref{sec:vortex} both analytically, using simple two--parameter
ansatz, and numerically, using micromagnetic simulations. The phase
diagram of possible ground states compares analytical predictions
and numerical simulations (Sec.~\ref{sec:phase}).

\section{The model. Homogeneous states}
\label{sec:model} %

We start with the continuum description of the spin distributions
inside the nanoring in terms of the  magnetisation, normalised by
its saturation value, $\vec{m} = \vec{M}/M_s$, and content ourselves
with two main contributions to the energy functional, the exchange
energy $E_{\text{ex}}$ and the magnetostatic energy $E_{\text{MS}}$.
We neglect the anisotropy energy contributions, what is a reasonable
for the soft materials like permalloy
($\text{Ni}_{80}\text{Fe}_{20}$, Py).

For the homogeneous magnetisation distribution the total energy
contains only the contribution of surface magnetostatic charges:
\begin{equation} \label{eq:E-MS-defn}
E_{\text{MS}} = \frac{1}{2} \int_S \int_{S'}
\frac{(\vec{M}\cdot\mathrm{d}\vec{S})(\vec{M'}\cdot\mathrm{d}\vec{S'})}{|\vec{r}
- \vec{r'}|}.
\end{equation}

Let us consider the nanoring, $R$ is the outer radius, $a$ is its
inner radius, and $h$ is the thickness. First of all we study the
case of the ring homogeneously magnetised along its plane. It is
convenient to consider the energy density, normalised by the value
$\pi R^2 h M_s^2$:
\begin{equation*} \label{eq:W-edge-polar}
W_{\text{MS}} = \frac{\varepsilon}{\pi} \int_S \int_{S'}
\frac{\rho\mathrm{d}\chi \mathrm{d}\zeta \rho'\mathrm{d}\chi'
\mathrm{d}\zeta'\; \cos\chi\cos\chi'}{\sqrt{\rho^2+\rho'^2-2\rho
\rho'\cos(\chi-\chi') + 4\varepsilon^2(\zeta - \zeta')^2}},
\end{equation*}
where $S$ indicates the outer and inner edge surfaces,
$(\rho,\chi,\zeta)$ are the cylindrical coordinates, $\rho$ is the
radius normalised by the magnitude of the outer ring radius $R$, the
variable $\zeta$ is the thickness normalised by $h$. We have
introduced also the particle \emph{aspect ratio} $\varepsilon =
h/2R$, and \emph{radii ratio} $\alpha = a/R$. Integration over edge
surface can be rewritten in explicit form:
\begin{subequations} \label{eq:W-homog}
\begin{align} \label{eq:W-edge}
W_{\text{MS}}^x &= \mathscr{E}(1,1) - 2\alpha \mathscr{E}(1,\alpha)
+
\alpha^2 \mathscr{E}(\alpha,\alpha),\\
\mathscr{E}(\rho,\rho') &= \frac{\varepsilon}{\pi}\!\!
\int\limits_0^{2\pi}\!\mathrm{d}\chi'\!
\int\limits_0^1\!\mathrm{d}\zeta\!
\int\limits_0^1\!\mathrm{d}\zeta'\!
\int\limits_0^{2\pi}\!\mathrm{d}\chi
\frac{\cos\chi\cos\chi'}{\sqrt{\rho^2+\rho'^2-2\rho
\rho'\cos(\chi-\chi') +4\varepsilon^2(\zeta - \zeta')^2}}. \nonumber
\end{align}
Straightforward but long and tedious calculations result
\begin{align}
\label{eq:E-rho&rho} %
\mathscr{E}(\rho,\rho) &= \frac{4\rho}{3\varepsilon}\left\{ -1 +
\frac1m\left[\frac{\varepsilon^2}{\rho^2} \text{K}(m) + \left(1-
\frac{\varepsilon^2}{\rho^2}\right)\text{E}(m) \right] \right\},\\
\label{eq:E-1&alpha} %
\mathscr{E}(1,\alpha) &=
\frac{\sqrt{1+\alpha^2}}{2\varepsilon}I(m_1) -
\frac{\sqrt{1+\alpha^2 + 4\varepsilon^2}}{2\varepsilon}I(m_2) +
\frac{4\varepsilon}{m_2\sqrt{\alpha}} \Bigl[\text{K}(m_2) -
\text{E}(m_2)\Bigr] \nonumber\\
&+ \frac{4\varepsilon m_2(1-m_1^2)}{m_1^2\sqrt{\alpha}}
\Bigl[\text{K}(m_2) - \Pi(m_1^2, m_2)\Bigr],\\
\label{eq:I(x)} %
I(x) &= \frac{8\sqrt{1-x^2/2}}{3x^2}\left[\frac{1-x^2}{1-x^2/2}
\text{K}(x)-\text{E}(x)\right],\\
m&=\frac{1}{\sqrt{1+\varepsilon^2/\rho^2}},\
m_1=\frac{2\sqrt{\alpha}}{1+\alpha}, \
m_2=\frac{2\sqrt{\alpha}}{\sqrt{(1+\alpha)^2 + 4\varepsilon^2}}.
\end{align}
\end{subequations}
Here $\text{K}(x),\;\text{E}(x)$ and $\Pi(x,y)$ are the complete
elliptic integrals of the first, second and third kind respectively.

In the limit case of the disk, the bulk expression
\eqref{eq:W-homog} transfers into the well--known \cite{Aharoni90}
formula
\begin{equation} \label{eq:E-disk}
W_{\text{MS}}^{x\text{ (disk)}} = \frac{4}{3\varepsilon}\left\{ -1 +
\frac1m\left[\varepsilon^2 \text{K}(m) + \left(1-
\varepsilon^2\right)\text{E}(m) \right] \right\}, \quad m =
\frac{1}{\sqrt{1+\varepsilon^2}}.
\end{equation}

To calculate the energy of the ring, homogeneously magnetised along
$z$--axis, one can use relations between the magnetometric
demagnetisation factors \cite{Akhiezer68}. These may be written in
the form:
\begin{equation} \label{eq:demagnetisation}
2W_{\text{MS}}^x+W_{\text{MS}}^z=2\pi(1-\alpha^2),
\end{equation}
where the factor $(1-\alpha^2)$ appears in consequence of our
normalisation of the energy density.

It is well known that the homogeneously in--plane magnetised
(easy--plane, EP) state is preferable energetically for thin enough
disk particles; when the disk aspect ratio is greater then the
critical value $\varepsilon_0\approx0.906$ \cite{Aharoni90}, the
homogeneously out--of--plane magnetised (easy--axis, EA) state is
realised.

\begin{figure}
\begin{center}
\includegraphics*[width=7.5cm]{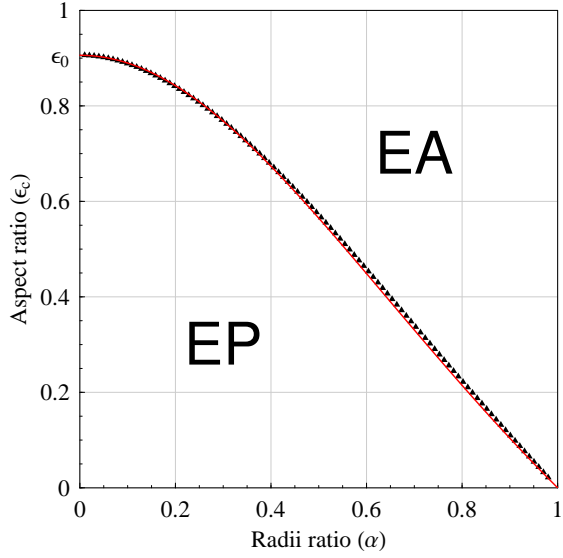}
\caption{The critical curve for the homogeneously magnetised states.
Symbols corresponds to the numerical solution (see the text), and
the curve is the approximate solution \eqref{eq:varepsilon-c-fit}.} %
\label{fig:epsCritVSalpha}
\end{center}
\end{figure}

Here we make a similar critical analysis for the nanoring. At
critical values of the parameters $W_{\text{MS}}^x =
W_{\text{MS}}^z$, and the critical curve $\varepsilon_c(\alpha)$ can
be found by solving the equation
$W_{\text{MS}}^x(\varepsilon_c,\alpha)=\frac{2\pi}{3}(1-\alpha^2)$.
The numerical solution of this equation is plotted in
Fig.~\ref{fig:epsCritVSalpha}. For the approximate description one
can use the asymptotically correct solution
\begin{equation} \label{eq:varepsilon-c-fit}
\varepsilon_c^{\text{approx}}(\alpha) = \varepsilon_0
\frac{1-\alpha^2}{1 - \alpha^2 + 2\alpha^2\varepsilon_0},
\end{equation}
which reproduces the numerical results with the accuracy within
$6\times 10^{-3}$, see Fig.~\ref{fig:epsCritVSalpha}.

\section{Vortex state}
\label{sec:vortex} %

The phase diagram, see Fig.~\ref{fig:epsCritVSalpha}, describes the
single--domain formation only. When the particle size exceeds
 the exchange length, $l_{\text{ex}} = \sqrt{A/4\pi
M_s^2}$ ($A$ is the exchange constant of the material), a
magnetisation curling takes place \cite{Hubert98}, and the
vortex--state can be energetically preferable. We start with the
exchange energy in the form $ E_{\text{ex}} =
\frac12Ah\int\mathrm{d}^2x(\vec{\nabla}\vec{m})^2$. Using the
angular parametrisation for the normalised magnetisation $\vec
m=\{\cos\theta\cos\phi;\cos\theta\sin\phi;\cos\theta\}$, one can
describe the vortex solution as follows:
\begin{equation} \label{eq:vortex-state}
\theta=\theta(r), \qquad \phi=\pm\frac\pi2+\chi.
\end{equation}
Here $(r,\chi)$ are the polar coordinates in the ring plane. For the
vortex--like solution \eqref{eq:vortex-state} the exchange energy
density takes a form:
\begin{equation} \label{eq:W-exchange}
W_{\text{ex}}^{\text{vortex}} = \frac{4\pi l_{\text{ex}}^2}{R^2}
\int_a^R r\mathrm{d}r\left[{\theta'(r)}^2 +
\frac{\sin^2\theta(r)}{r^2}\right].
\end{equation}
Let us calculate the magnetostatic energy. For the vortex
distribution \eqref{eq:vortex-state} the volume magnetostatic
charges are absent ($\vec{\nabla}\cdot\vec{m}=0$), and the surface
contribution \eqref{eq:E-MS-defn} can be presented in the form
\cite{Hubert98}
\begin{equation} \label{eq:W-MS-z}
W_{\text{MS}}^{\text{vortex}} = \frac{2\pi}{\varepsilon}
\int_0^\infty \mathrm{d}x  \left(1-e^{-2\varepsilon x}\right)
\left[\int_\alpha^1 \rho \mathrm{d} \rho \cos\theta(\rho) J_0(\rho
x)\right]^2.
\end{equation}

At this point we should specify the profile function $\theta(\rho)$.
There are known several models for describing the vortex structure
in the disk--shape particles. We should mention the core model by
\citet{Usov93} and its modifications
\cite{Metlov02,Scholz03,Landeros05}, which neglects the
magnetostatic interaction inside the vortex core, and also has a
singularity at the core radius. More realistic model was proposed by
\citet{Hoellinger03}, where the vortex has a bell--shaped structure.
This model is in a good agreement with simulations data for a disk.
Nevertheless we will use another model by the following reasons: (i)
the model should be able to describe magnetisation distributions
both in disks and rings, and (ii) it should be very simple to be
treated analytically. All mentioned above models can not be simply
generalised for the ring case, and can not be analysed analytically.

In pursuance of above mentioned reasons we propose  the following
\emph{two--parameters ansatz}:
\begin{equation} \label{eq:2param-ansatz}
\cos\theta(r) = \mu \exp\left[-\left(\frac{r}{\lambda
l_{\text{ex}}}\right)^2\right].
\end{equation}
Here parameter $\mu$ describes the vortex amplitude in the centre of
the ring. The typical vortex width is determined by the parameter
$\lambda$. An advantage of this ansatz is a possibility to describe
different kinds of vortices: the OPV for $\mu=1$, the IPV for
$\mu=0$, and the IMV for $\mu\in(0,1)$. For the case of the disk
($\mu=1$) our ansatz \eqref{eq:2param-ansatz} fits well the
structure of the pure OPV in easy--plane magnets \cite{Mertens00}.

For the analytical treatment of the model, we make also one serious
simplification: we use the local shape--anisotropy model
\cite{Sheka05} instead of the nonlocal magnetostatic energy
\eqref{eq:W-MS-z}, what is acceptably for thin particles
($\varepsilon \ll1$):
\begin{equation}
\label{eq:WmsVortex}%
W_{\text{anis}} = \frac{4\pi}{R^2}\int_a^R r\mathrm{d}
r\cos^2\theta(r).
\end{equation}

Finally the total energy density for the vortex state ring reads
\begin{equation} \label{eq:W4ring}
W_{\text{IMV}} = \frac{4\pi l_{\text{ex}}^2}{R^2}
\int\limits_{\frac{a}{\lambda l_{\text{ex}}}}^{\frac{R}{\lambda
l_{\text{ex}}}} x\mathrm{d}x \left[ \frac{1}{x^2} + \mu^2
e^{-2x^2}\left( \frac{4x^2}{1-\mu^2 e^{-2x^2}} - \frac{1}{x^2} +
\lambda^2 \right) \right].
\end{equation}
This integral can be derived analytically, the explicit form is
calculated in Appendix \ref{sec:appendix}. We have derived the
energy \eqref{eq:W4ring} using the local shape--anisotropical model
\eqref{eq:WmsVortex} instead of the non--local magnetostatic one
\eqref{eq:W-MS-z}. Strictly speaking, the simplified model is valid
only for infinitesimally thin rings, see Appendix
\ref{sec:appendix-MS} for details. However further analysis
demonstrates that it provides a reasonably accurate picture of
phenomenon even for a large aspect ratio.

\subsection{Pure out--of--plane vortex}

\begin{figure}
\begin{center}
\includegraphics[width=7.5cm]{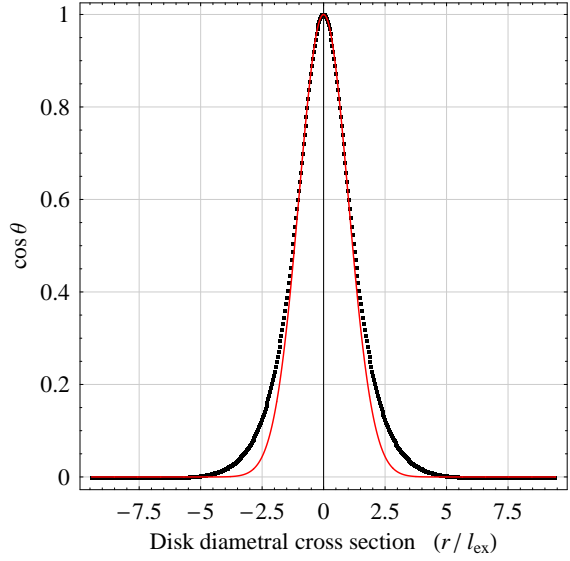}
\caption{The comparison of two OPV profiles : the ansatz
\eqref{eq:2param-ansatz} with $\mu=1$ and $\lambda=\sqrt2$ (solid
red line) and micromagnetic simulations (black bars) for the Py disk
($R=50$nm, $h=5$nm).} \label{fig:diskTheoryVSoommf}
\end{center}
\end{figure}

We start our analysis with the case of a disk--shape nanoparticle,
where the pure OPV is realised. This well--known case is a good test
for our simple theory. The magnetisation in the vortex centre is
perpendicular to the disk plane, therefore the vortex amplitude at
origin is equal to unit, $\mu=1$.

The energy of the vortex state nanodisk can be easily derived as a
limit case of Eq.~\eqref{eq:W4ring}, see Appendix \ref{sec:appendix}
for details:
\begin{equation} \label{eq:W-OPV}
\begin{split}
W_{\text{OPV}}=\cfrac{2\pi
l_{\text{ex}}^2}{R^2}\Biggl[&\gamma-\frac12\xi_R^2-\mathrm{Ei}(-\xi_R)+\ln\xi_R
+\xi_R\ln\left(e^{\xi_R}-1\right)\\
&-\int\limits_1^{e^{\xi_R}}\cfrac{\ln(t-1)}{t} \mathrm{d}t\Biggr]
+2\pi\cfrac{1-e^{-{\xi_R}}}{\xi_R}.
\end{split}
\end{equation}
Here $\xi_R=2R^2/\lambda^2l_{\text{ex}}^2$, $\gamma\simeq0.577$ is
Euler's constant; $\mathrm{Ei}(\xi)$ is the exponential integral
function. The value of the variational parameter $\lambda$ can be
found by the minimisation of the energy \eqref{eq:W-OPV} with
respect to $\lambda$. Finally, the $\lambda$--parameter can be
calculated as a solution of transcendental equation
\begin{equation} \label{eq:lambda}
\lambda^2(R)=2\cfrac{1+e^{-2\xi_R}+e^{-\xi_R}(\xi_R^2-2)}{1+e^{-2\xi_R}(1+\xi_R)-e^{-\xi_R}(\xi_R+2)}\;,
\end{equation}
which can be solved numerically. For the case of large enough disk
radii ($R>4l_{\text{ex}}$), the numerical solution $\lambda(R)$
practically coincides (with the accuracy within $2\times 10^{-5}$)
with the limit value $\lambda(\infty) = \sqrt{2}$.

To verify our analytical  model calculations we have preformed the
numerical computer simulations using a three--dimensional OOMMF
micromagnetic simulator code \cite{OOMMF}. In all simulations we
have used the following material parameters for the Py:
$A=1.3\times10^{-6}$ erg/cm (using SI units $A^{\text{SI}}=1.3\times
10^{-11}$ J/m), $M_s=8.6\times10^{2}$ G  ($M_s^{\text{SI}}=8.6\times
10^5$ A/m), and the anisotropy have been neglected. This corresponds
to the exchange length $l_{\text{ex}} = \sqrt{A/4\pi M_s^2}\approx
5.3$nm ($l_{\text{ex}}^{\text{SI}} = \sqrt{A/\mu_0 M_s^2}$). A
comparison of the vortex profiles, obtained by simulations and by
analytical approach is presented in
Fig.~\ref{fig:diskTheoryVSoommf}. Our ansatz
\eqref{eq:2param-ansatz} fits the simulation data with the accuracy
within $0.09$.

\subsection{Intermediate vortex solution}

\begin{figure}
\begin{center}
\includegraphics[width=7.5cm]{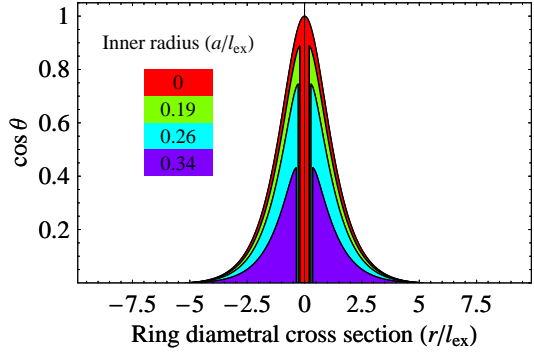}
\caption{Micromagnetic simulation data of the IMV profiles for
different radii of inner hole. Other parameters are the same as in
Fig.~\ref{fig:diskTheoryVSoommf}.}
\label{fig:allShapes} %
\end{center}
\end{figure}

Let us consider the ring--shape nanoparticle. The presence of the
hole in the centre of the disk changes the topological properties of
the vortex solution. Hence it is not necessary for the vortex
amplitude to be equal to unit in the ring centre. Moreover, if the
inner hole is big enough, there can exist pure IPV, where the
magnetisation does not have out--of--plane component at all. In the
intermediate case of small enough inner hole the vortex amplitude
$\mu$ should vary in the range $[0,1]$. We have verified this idea
by the micromagnetic simulations, as described above. One can see
from Fig.~\ref{fig:allShapes} that the out--of--plane vortex
structure has a well--defined bell shape, but the amplitude of the
vortex decays fast when the inner hole becomes bigger. When the
inner hole is greater than some critical value $a_c$, which is about
$2$nm for the Py ring, only pure IPV solutions can be realised.

\begin{figure}
\begin{center}
\includegraphics*[width=7.5cm]{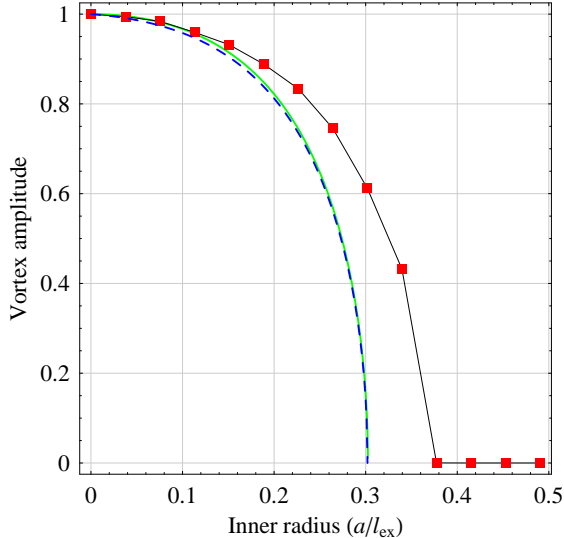}
\caption{Vortex amplitude (at the inner edge) vs inner radius of the
ring. Bars denotes simulations results and solid green line
corresponds to theoretical calculation, which were performed
numerically. Dashed blue line corresponds to the fit
(\ref{eq:mu(a)-fit}). Other parameters are the same as in
Fig.~\ref{fig:diskTheoryVSoommf}. }
\label{fig:vAmpl} %
\end{center}
\end{figure}

We have chosen the ansatz \eqref{eq:2param-ansatz} because the bell
shapes of the vortex structures are well pronounced for different
vortex amplitudes. Using this ansatz we have derived the general
expression for the vortex energy \eqref{eq:W4ring}. To calculate
variational parameters of our model $\mu$ and $\lambda$, we have
minimised the energy \eqref{eq:2param-ansatz} with respect to these
parameters: $\partial W/\partial \mu = 0$ and $\partial W/\partial
\lambda = 0$. This set of equations can be solved only numerically.
Analysis shows that the solution practically (with the accuracy
within $1\times 10^{-4}$) does not depend on $R$ when $R>3
l_{\text{ex}}$, hence one can use the limit case $R\to\infty$ in
calculations. The results are presented in Fig.~\ref{fig:vAmpl}.
Both analytical and simulation results have the same main behaviour:
the vortex amplitude slightly decays when the inner hole increases
for small enough holes, but it sharply breaks at some critical inner
radius $a_c$. However, the value of the critical inner radius is
about $2$~nm from simulations while $a_c=1.6$~nm from the theory
(see below). This discrepancy in critical values is caused by the
local model of the magnetostatic interaction. In Appendix
\ref{sec:appendix-MS} we will take into account the nonlocal
magnetostatic interaction, which gives the critical inner radius
1.94 nm, what is in a good agreement with simulations results.

\begin{figure}
\begin{center}
\includegraphics[width=7.5cm]{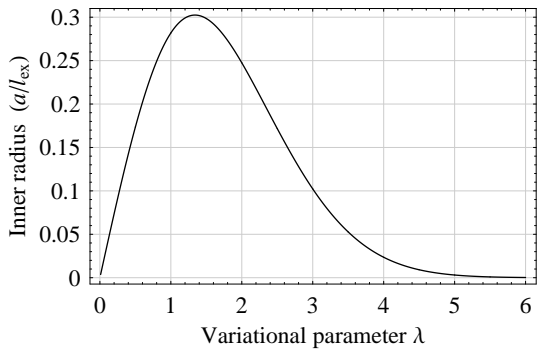}
\caption{The numerical solution of the Eq.~\eqref{eq:a-cr}.}
\label{fig:acrVSlambda}
\end{center}
\end{figure}

To calculate the critical value $a_c$, when the transition to IPV
occurs, one can expand the IMV energy density \eqref{eq:W4ring} in
series on $\mu$ at $\mu=0$:
\begin{equation} \label{eq:I-series}
\begin{split}
W_{\text{IMV}} &\approx \frac{\mu^2}{2}C_1(\lambda) +
\frac{\mu^4}{4} C_2(\lambda),\\
C_1(\lambda) &= \frac{8\pi
l_{\text{ex}}^2}{R^2}\int_{\frac{a}{\lambda
l_{\text{ex}}}}^\infty\mathrm{d}x\ xe^{-2x^2}\left( 4x^2 -
\frac{1}{x^2}
+ \lambda^2\right),\\
C_2(\lambda) &= \frac{64\pi l_{\text{ex}}^2}{R^2}
\int_{\frac{a}{\lambda l_{\text{ex}}}}^\infty \mathrm{d}x\
x^3e^{-4x^2}.
\end{split}
\end{equation}
One can see that $C_2$ is always positive, hence the vortex solution
with $\mu\neq0$ can exist only when $C_1<0$. This is the case of the
double--well potential with minima at $\mu^2= |C_1|/C_2$. The pure
IPV solution with $\mu=0$ corresponds to the case of $C_1>0$. Thus
the critical point can be found from the condition $C_1=0$. The
numerical solution of the equation
\begin{equation} \label{eq:a-cr}
\int_{\frac{a}{\lambda l_{\text{ex}}}}^\infty \mathrm{d}x\
xe^{-2x^2}\left( 4x^2 - \frac{1}{x^2} + \lambda^2\right)=0
\end{equation}
is shown in Fig.~\ref{fig:acrVSlambda}. It is not necessary to check
the second minimum condition with respect to $\lambda$. Obviously,
the critical parameters correspond to the maximum of the curve
$a(\lambda)$. Finally, the critical inner radius $a_c \approx
0.3l_\text{ex}$, see Appendix \ref{sec:appendix}.

At vicinity of the critical point, the vortex amplitude has
asymptotic $\mu\propto\sqrt{a_c-a}$. One can extend this result and
use the function
\begin{equation} \label{eq:mu(a)-fit}
\mu^{\text{approx}} = \sqrt{\frac{1-a/a_c}{(1-a/l_{\text{ex}})^3}},
\end{equation}
which fits the numerical results of Fig.~\ref{fig:vAmpl} with high
accuracy (about $9\times10^{-3}$) in the whole range of the
parameters.

When the inner ring radius $a$ exceeds the critical value $a_c$, the
pure IPV solution with $\mu=0$ takes place. In this case all
magnetostatic charges are absent, and the vortex energy density has
only exchange contribution, which simply reads
\begin{equation}
\label{eq:IPV}%
W_{\text{IPV}} = \cfrac{4\pi l_\mathrm{ex}^2}{R^2}\ln\frac Ra.
\end{equation}

\section{Phase diagrams}
\label{sec:phase} %

\begin{figure}
\begin{center}
\includegraphics[width=6.89cm]{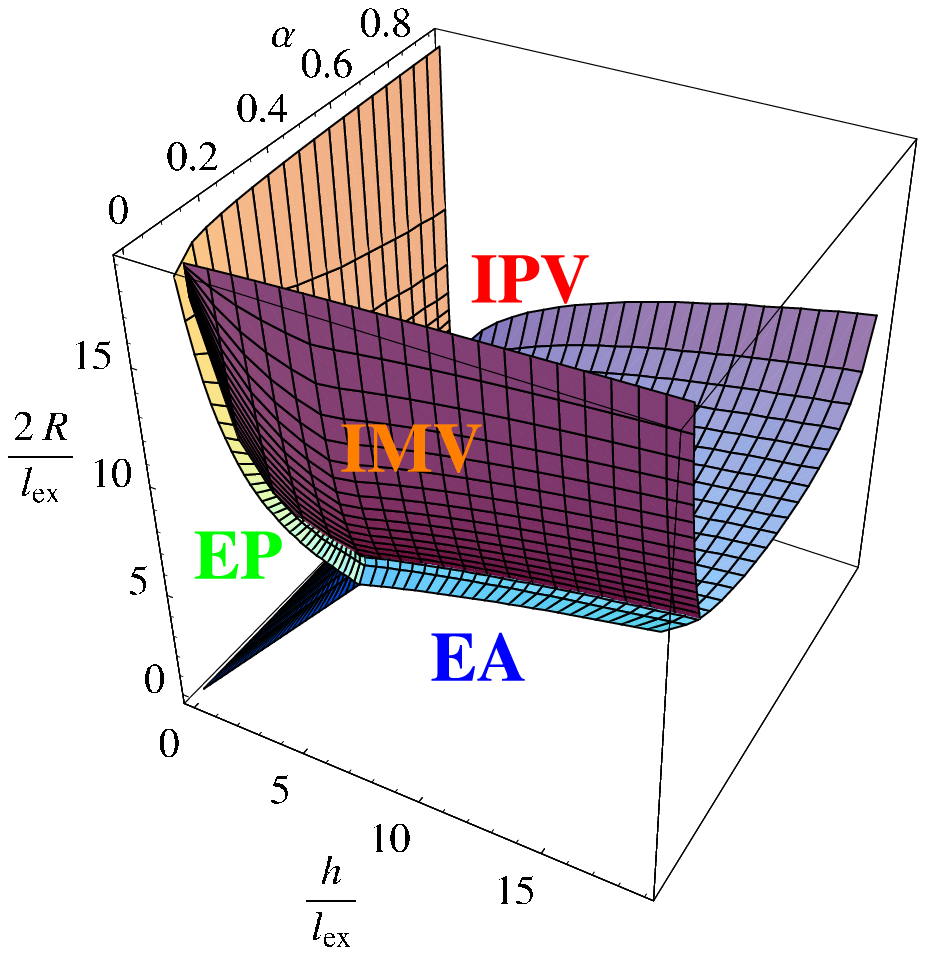}
\includegraphics[width=6.89cm]{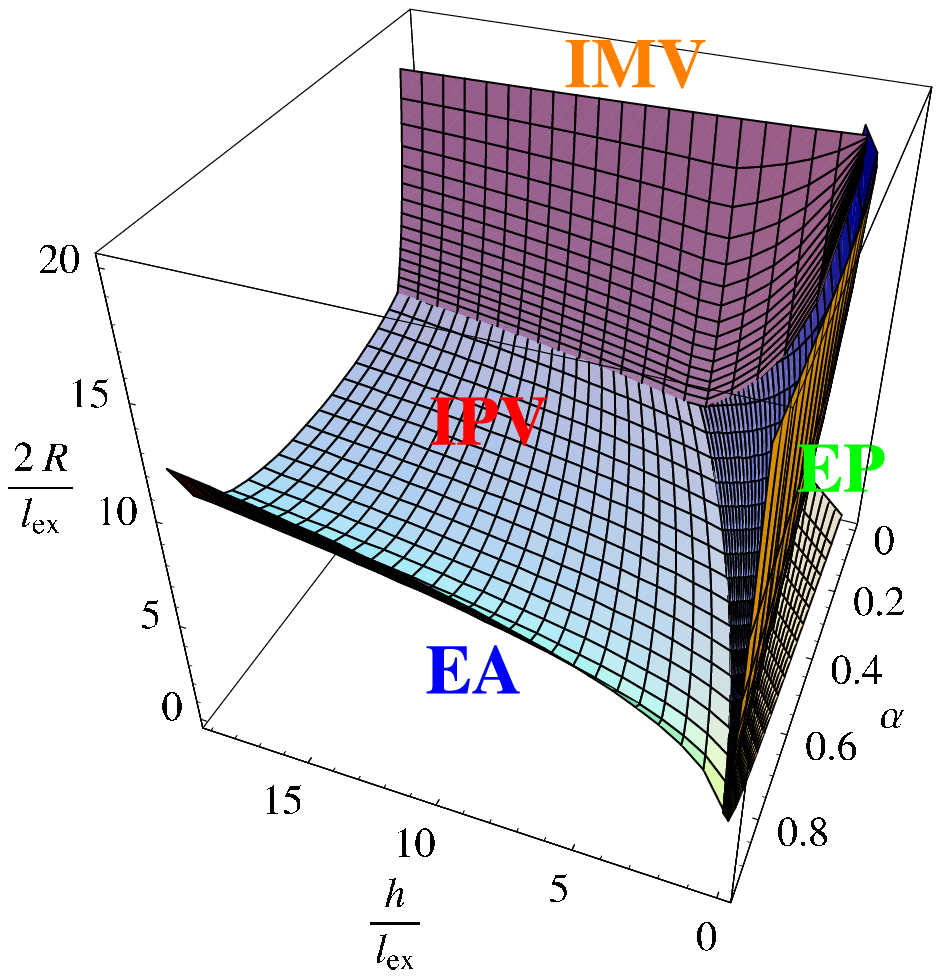}
\caption{Three--dimensional phase diagram of magnetisation ground
states from analytical analysis.} \label{fig:3D-PD}
\end{center}
\end{figure}

Summarising results on the equilibrium magnetisation distribution,
we have calculated energetically preferable states for different
ring geometries. Since we have three parameters, which define the
ring,  $R$, $a$, an $h$, our phase diagram is a three--dimensional
plot. Theoretically calculated phase diagram is presented in
Fig.~\ref{fig:3D-PD}. Different phases are separated by boundary
surfaces. The general properties of the phase diagram are as
follows. The ground state of the very thin ring is the homogeneous
easy--plane state. At the thickness increasing we can switch either
to the homogeneous easy--axis state, when the outer ring radius is
small enough, or to the vortex state, when the ring is large. The
boundary surface between two homogeneous states can be simply
expressed analytically, $h=2R\varepsilon_c(\alpha)$, where the
$\varepsilon_c(\alpha)$--dependence is described by
Eq.~\eqref{eq:varepsilon-c-fit}, see also
Fig.~\ref{fig:epsCritVSalpha}. Now, if we increase the inner radius
of the ring, the vortex state becomes energetically preferable. One
can follow step--by--step transitions between different vortex
phases: from the pure OPV state for the case of the disk ($a=0$) to
the IMV state for the ring with small radii ratio ($a<a_c$) and,
finally, to the pure IPV state when $a>a_c$.

\begin{figure}
\begin{center}
\subfloat[$\alpha=0$]{\label{fig:2D-PD-alpha=0}%
\includegraphics[width=6.89cm]{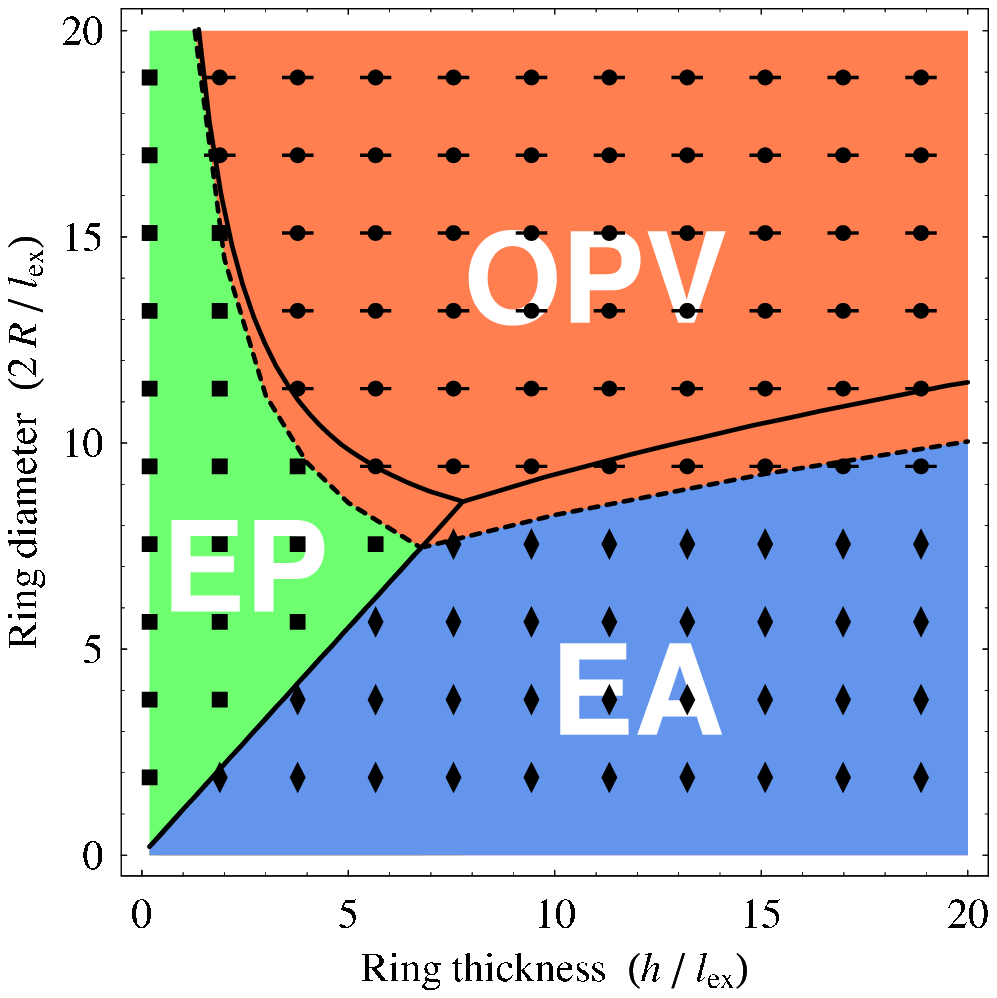}}
\subfloat[$\alpha=0.1$]{\label{fig:2D-PD-alpha=0.1}%
\includegraphics[width=6.89cm]{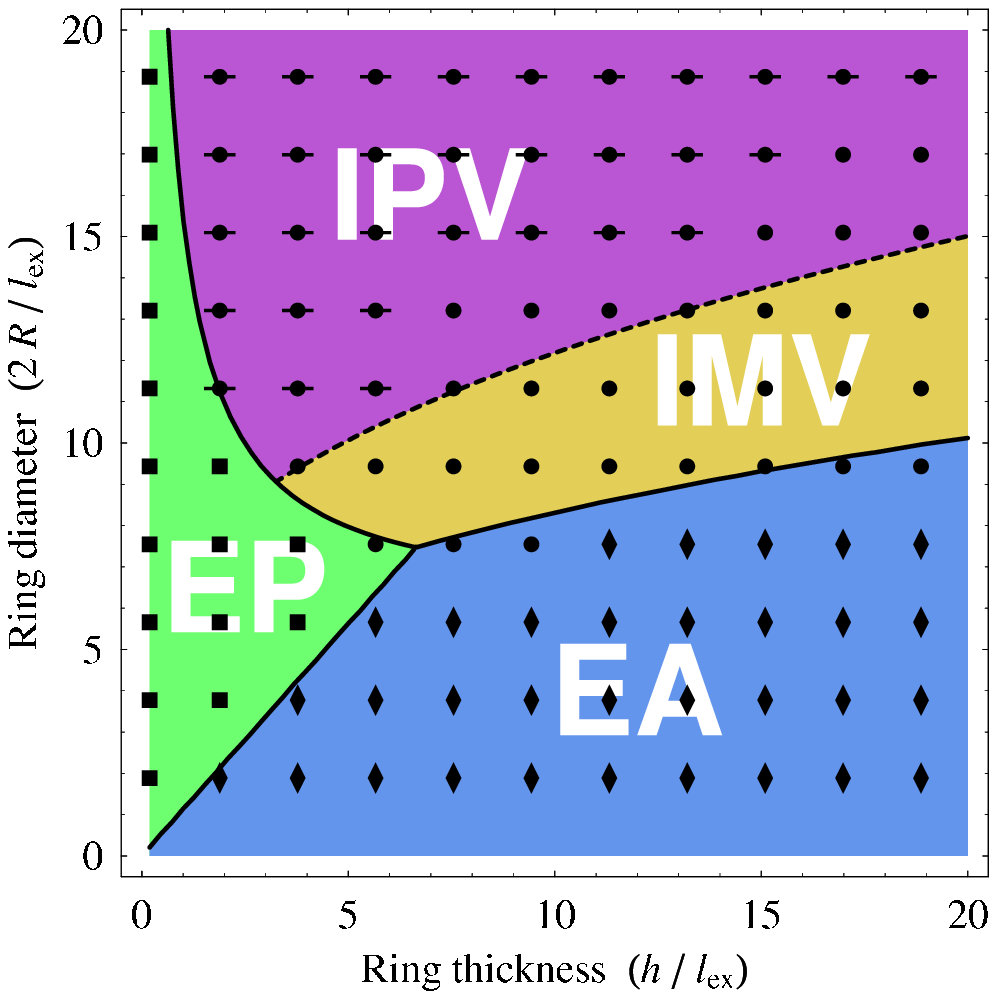}}\\
\subfloat[$\alpha=0.25$]{\label{fig:2D-PD-alpha=0.25}%
\includegraphics[width=6.89cm]{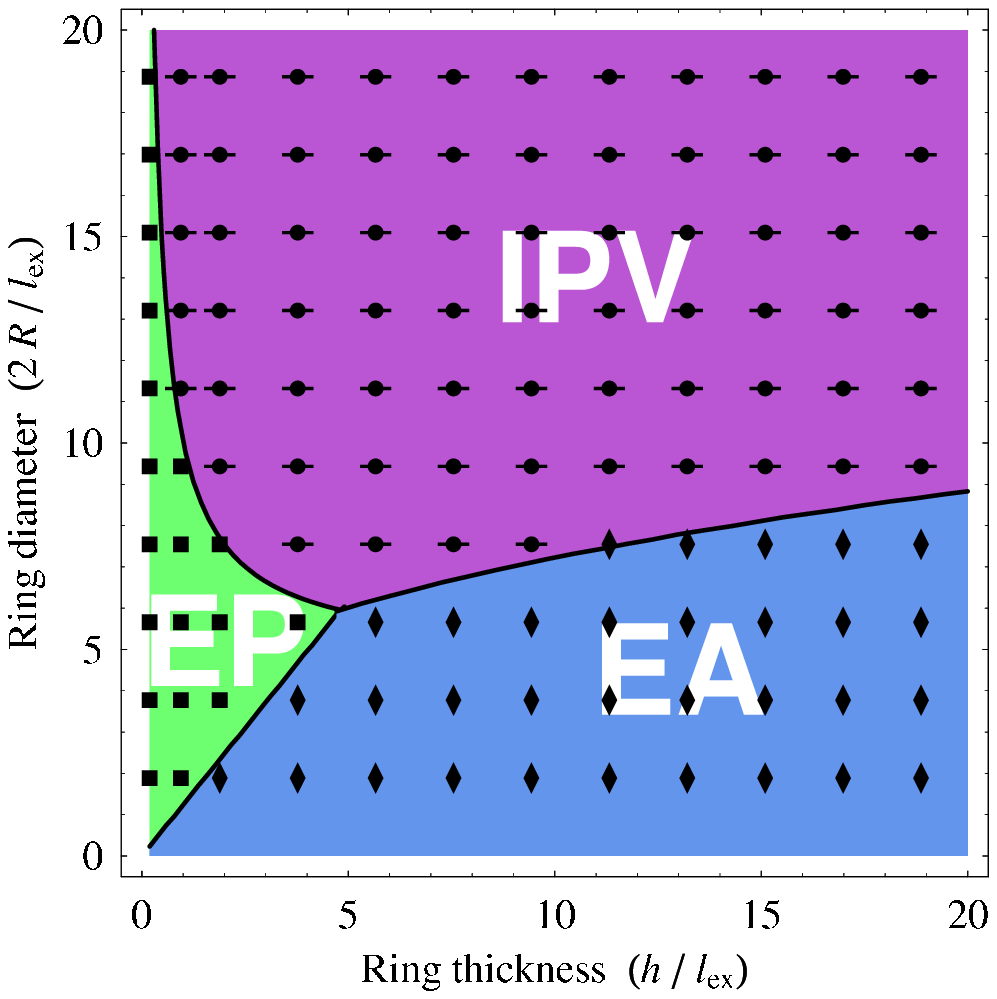}}
\subfloat[$\alpha=0.5$]{\label{fig:2D-PD-alpha=0.5}%
\includegraphics[width=6.89cm]{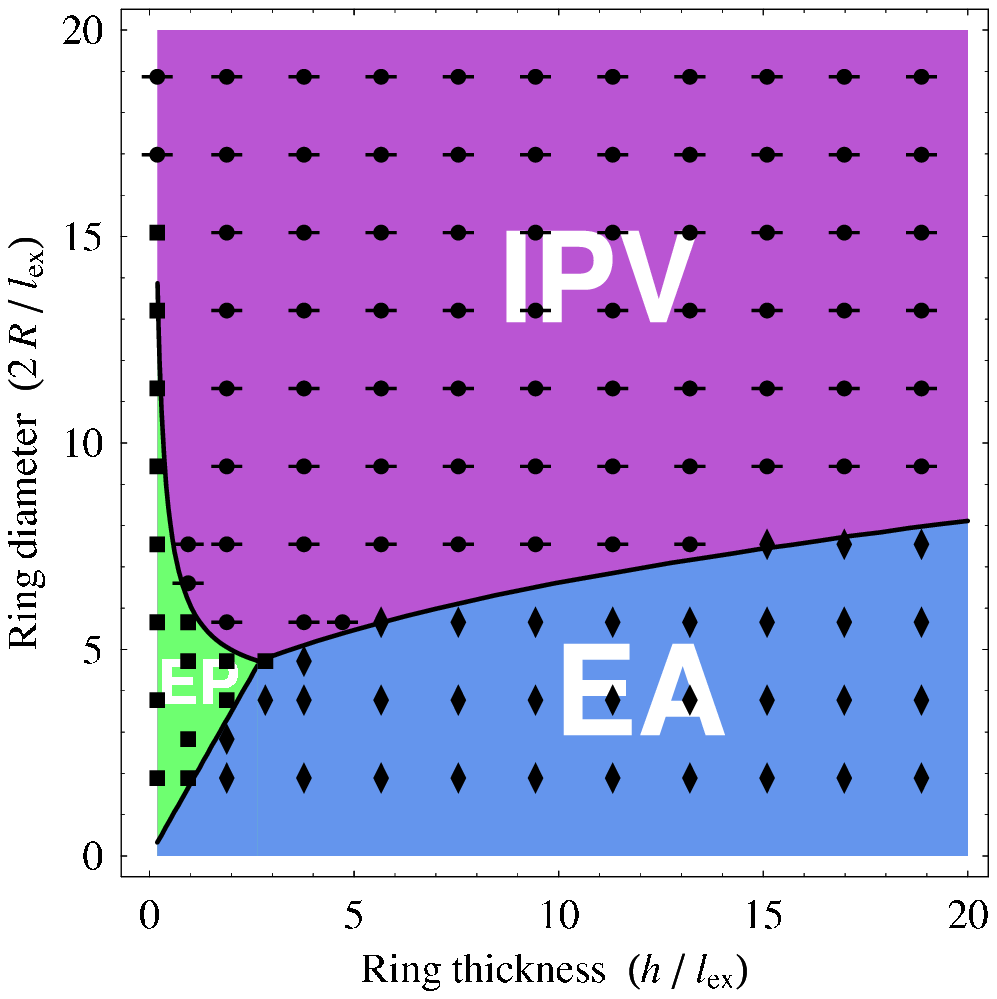}}
\caption{The phase diagram of magnetic ground states for different
radii ratio. The areas of different colours correspond to different
kinds of magnetisation distribution. Symbols represent the
simulation data: squares -- EP, diamonds -- EA, disks -- IMV, disks
with vertical lines -- OPV, disks with horizontal lines -- IPV. The
dashed lines in fig. (a) and (b) are the theoretical borders
calculated within the nonlocal model of magnetostatic interaction.
Solid lines correspond to the theoretical borders, calculated within
the local model.} \label{fig:2D-PD}
\end{center}
\end{figure}

To verify our theoretically calculated phase diagram, we use 3D
micromagnetic simulations, as described above. Simulations data are
reproduced in Fig.~\ref{fig:2D-PD} together with theoretical
results. Fig.~\ref{fig:2D-PD-alpha=0} presents the phase diagram for
the case of the nanodisk. Numerical results for this case is in a
good agreement with previous theoretical \cite{Metlov02,Scholz03}
and experimental \cite{Ross02} results. One can compare the
theoretical results obtained within the local (solid line) and
nonlocal (dashed line) models of the magnetostatic interaction. In
the case of thin disks the boundaries are coincident and agree with
simulations results. However in the case of thick disks for the
boundary between OPV phase and homogeneous EA solid line (the local
model) lies higher then simulations data. This is because the
variational approach in the local shape--anisotropy model always
gives the upper limit of the energy.

The presence of the hole inside the disk drastically changes the
vortex state. The vortex amplitude depends on the inner radius of
the ring. If the radii ratio is small enough, the phase diagram
contains four phases, see Fig.~\ref{fig:2D-PD-alpha=0.1}. The
boundary between phases IMV and IPV was calculated using nonlocal
model of magnetostatic interaction (see Appendix
\ref{sec:appendix-MS}). We have not managed to calculate the
boundaries between homogenous and vortex states within nonlocal
model, because in the case of ring there are two variational
parameters ($\mu$ and $\lambda$), in contrast to the case of disk,
where only one variational parameter ($\lambda$) is present. If the
inner radius exceeds the critical value $a_c$, only one vortex phase
with IPVs can exist, see Fig.~\ref{fig:2D-PD-alpha=0.25} and
Fig.~\ref{fig:2D-PD-alpha=0.5}.

\begin{figure}
\begin{center}
\subfloat[]{\label{fig:TriplePoint-R-h}%
\includegraphics[width=6.89cm]{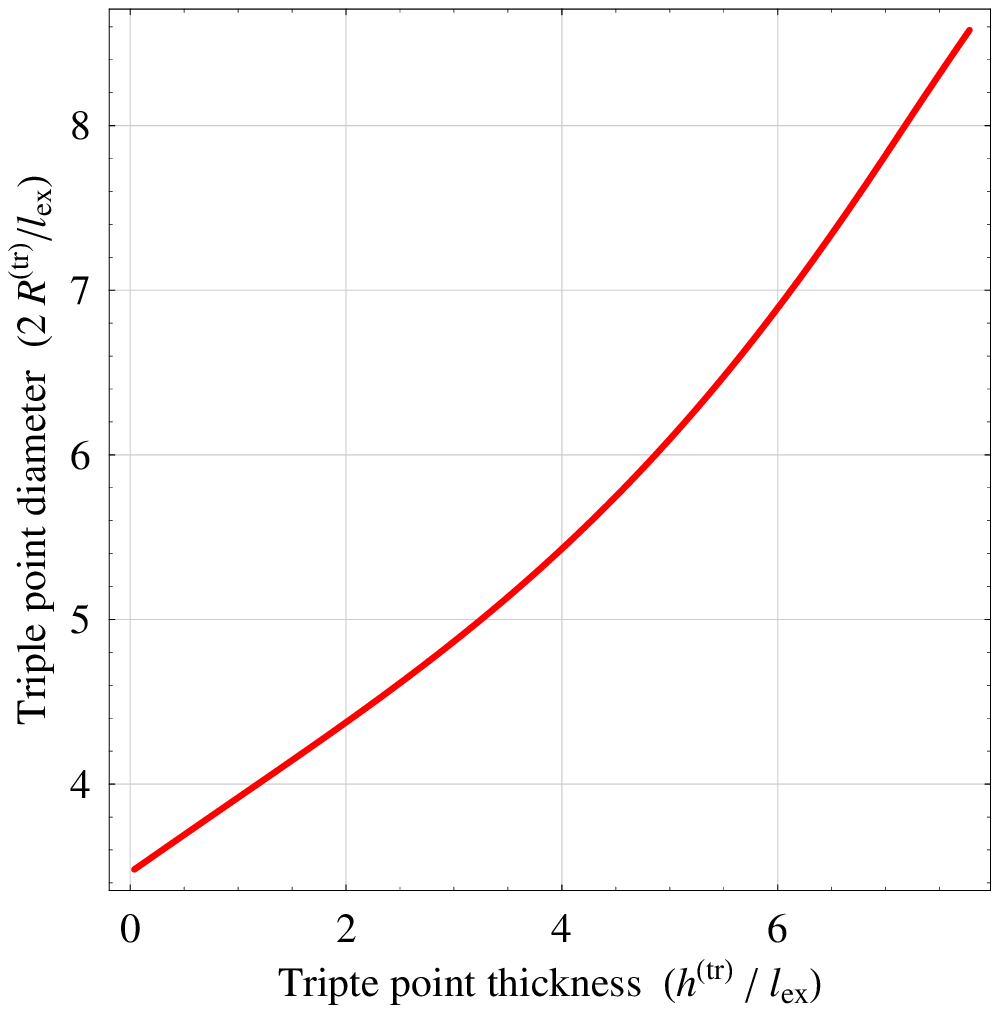}}
\subfloat[]{\label{fig:TriplePoint-R-alpha}%
\includegraphics[width=6.89cm]{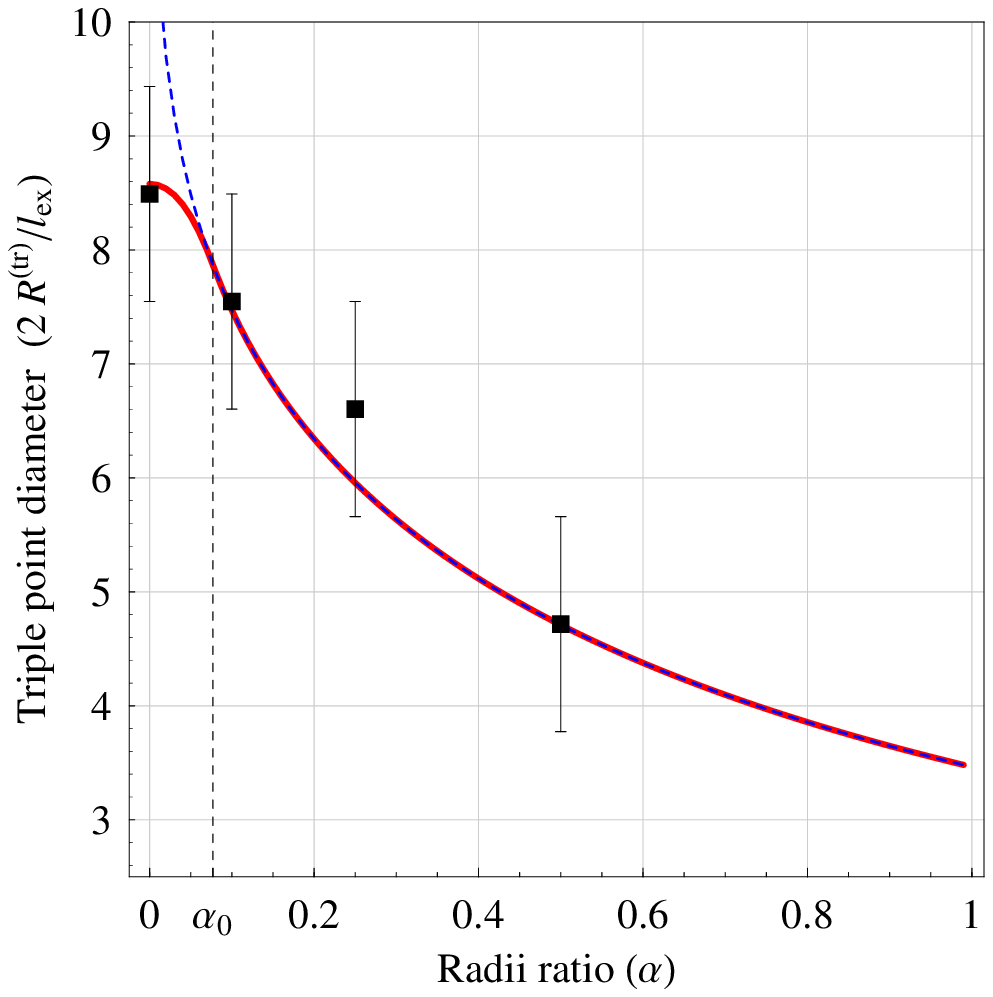}}
\caption{The triple point analysis: the outer ring diameter as a
function of (a) the ring thickness and (b) the radii ratio. Red
lines corresponds to the exact numerically calculated data (the
local model of the magnetostatics), dashed blue line is the
analytical solution \eqref{eq:triple4IPV}, which is acceptable for
$\alpha>\alpha_0$. Points with error bars were obtained from
simulations data (Fig.~\ref{fig:2D-PD})}
\label{fig:TriplePoint} %
\end{center}
\end{figure}

The \emph{triple} point in the phase diagrams is the object of a
special interest. In particular, it provides an information about
the lowest possible radius of the vortex state nanoparticle. The
triple point was analysed for the disk particles using the scaling
technique \cite{Castro02,Landeros05}. Here we present results for
the case of the ring. Since at this point the vortex energy
$W_{\text{IMV}}$ is equal to the homogeneous state energies,
moreover $W_{\text{MS}}^x = W_{\text{MS}}^z $, one can calculate the
triple point parameters by solving an equation $W_{\text{IMV}} =
\frac{2\pi}{3}(1-\alpha^2)$. In general case it can be done
numerically only, see Fig.~\ref{fig:TriplePoint}. An analytical
analysis can be performed asymptotically for small $\alpha$, where
we can limit ourselves to the linear dependence $2R^{\text{(tr)}} =
h^{\text{(tr)}}/\varepsilon_0$, where $\varepsilon_0\approx 0.906$
is the critical aspect ratio for a disk \cite{Aharoni90}. In the
limit case of $\alpha=0$ the triple point is characterised by the
following parameters: $R^{\text{(tr)}} \approx 3.73l_{\text{ex}}$
and $h^{\text{(tr)}} \approx 6.76 l_{\text{ex}}$ (The values were
obtained for the nonlocal model of the magnetostatics). By
increasing the inner radius, the triple point radius
$R^{\text{(tr)}}$ decreases, so the IMV can exist in smaller
particles than the pure OPV. At some critical inner radius $a_c$ the
transition to the IPV state occurs. The energy of the pure IPV has a
simple form \eqref{eq:IPV}, so the triple point analysis can be
easily done for this case:
\begin{equation}
\label{eq:triple4IPV}%
R^{\text{(tr)}}(\alpha)=l_{\textrm{ex}}\sqrt{\cfrac{6}{1-\alpha^2}
\ln\frac1\alpha},\qquad h^{\text{(tr)}}(\alpha) =
2R^{\text{(tr)}}(\alpha)\varepsilon_c(\alpha),\quad \alpha >
\alpha_0,
\end{equation}
where $\varepsilon_c(\alpha)$--dependence is described by
Eq.~\eqref{eq:varepsilon-c-fit}. Such simple analytical description
of the triple point is valid only, when $a>a_c$ and the IPV is
stable. Simple calculations show that it can be realised for
$\alpha>\alpha_0\approx 0.077$. The triple point radius
$R^{\text{(tr)}}$ and thickness $h^{\text{(tr)}}$ diminish when
$\alpha$ increases, and they have the asymptotic behaviour:
\begin{equation}
\label{eq:triple4IPV-as}%
R^{\text{(tr)}}(\alpha) \underset{\alpha \to 1}\sim
l_{\text{ex}}\sqrt3 \bigl[1 + (1-\alpha)/2\bigr], \qquad
h^{\text{(tr)}}(\alpha) \underset{\alpha \to 1}\sim
2l_{\text{ex}}\sqrt3(1-\alpha).
\end{equation}
The limit value $R^{\text{(tr)}}(1)=l_{\text{ex}}\sqrt3$ provides
the lowest bound for the vortex state magnetic ring. For the Py
nanoring this limit is about $9.18$nm.

\section{Conclusions}

We have presented a detailed study of the ground state of magnetic
nanorings, including homogeneous (easy--axis and easy--plane) states
and inhomogeneous vortex states. In addition to conventional
out--of--plane and in--plane vortex types we have found a new type
of vortices in a nanoring. Such intermediate vortices are
characterised by the smaller amplitude in the vortex centre, they
are the key point to understand the transition from OPV to IPV
state. We have proposed also a simple analytical description of the
IMV state. Using this approach we have studied analytically phase
transitions between different ground states. All results are
confirmed by our direct 3D micromagnetic simulations.

Using the critical analysis of the phase diagram triple point, we
conclude that the lower bound of the vortex--state nanoparticle
decreases for a ring geometry in comparison with a disk. The minimal
nanoring radius varies from about $3.7l_{\text{ex}}$ for the case of
the disk to $\sqrt3l_{\text{ex}}$ for the case of narrow ring.

Our model of the IMV can be applied for the description of dynamical
effects of the switching of the vortex polarisation
\cite{Gaididei00,Pulwey01,Zagorodny03}. We expect that the dynamical
switching effects should be more pronounced in the ring geometry,
because the vortex amplitude $\mu$ can smoothly vary and even change
its sign. This is the subject of future research.

\section{Acknowledgments}

Authors acknowledge the support from Deutsches Zentrum f{\"u}r Luft-
und Raumfart e.V., Internationales B{\"u}ro des BMBF in the frame of
a bilateral scientific cooperation between Ukraine and Germany,
project No. UKR 05/055. All simulations results presented in the
work were obtained using the computing cluster of Kiev University
\cite{unicc}. V.~Kr. thanks O.~Sudakov for productive consultations
about the computing cluster using. D.~Sh. thanks the University of
Bayreuth, where part of this work was performed, for kind
hospitality and acknowledges support from the Alexander von Humboldt
Foundation.

\appendix

\setcounter{equation}{0}
\renewcommand{\theequation}{\thesection.\arabic{equation}}
\makeatletter
\def\@seccntformat#1{\csname Pref@#1\endcsname \csname the#1\endcsname\quad}
\def\Pref@section{Appendix~}
\makeatother

\section{The vortex state energy calculations}
\label{sec:appendix}

Let us start with the vortex energy in the form \eqref{eq:W4ring}.
Using the direct integration one can write down the energy in the
form
\begin{equation} \label{eq:W-calculation}
W_\mathrm{IMV} = \frac{2\pi {l_{\text{ex}}^2}}{R^2}
\Bigl[\mathcal{I}(\xi_a)- \mathcal{I}(\xi_R)\Bigr].
\end{equation}
Here $\xi_a = \dfrac{2a^2}{l_{\text{ex}}^2 \lambda^2}$ and $\xi_R =
\dfrac{2R^2}{l_{\text{ex}}^2 \lambda^2}$; the integral $\mathcal
I(\xi)$ can be expressed as follows
\begin{equation}
\begin{split} \label{eq:I1}
\mathcal I(\xi)=&\frac{\xi^2}{2}+\frac{\lambda^2\mu^2}{2}e^{-\xi}+\mu^2\mathrm{Ei}(-\xi)+\int\limits_1^{e^\xi/\mu^2}\frac{\ln(t-1)}{t}\mathrm{d}t-\\
-&\ln\xi- \xi\ln\left(\frac{e^\xi}{\mu^2}-1\right).
\end{split}
\end{equation}
where $\mathrm{Ei}(x)$ is the exponential integral function. %

Using the asymptotical behaviour
\begin{equation} \label{eq:as4I}
\begin{split}
\mathcal{I}(\xi ) &\underset{\xi  \to 0}{\sim}
\frac{\lambda^2\mu^2}{2}+\mu^2\gamma-\ln\xi(1-\mu^2)+
\int\limits_1^{1/\mu^2}\frac{\ln(t-1)}{t}\mathrm{d}t,
\end{split}
\end{equation}
one can easy calculate that $\mathcal{I}(0)={\lambda^2}/{2}+\gamma$
when $\mu =1$. According to \eqref{eq:W-calculation} the vortex
energy of a disk with radius $R$ can be written down as
\begin{equation}\label{eq:disk-vortex-energy}
W_\mathrm{OPV} = \frac{2\pi {l_{\text{ex}}^2}}{R^2}
\Bigl[\mathcal{I}(0)- \mathcal{I}(\xi_R)\Bigr]_{\mu=1},
\end{equation}
which results in \eqref{eq:W-OPV}.

Let us go back to the case of the ring and calculate the critical
inner radius, when the transition from the intermediate vortex
solution to the pure in--plane one takes place. One can rewrite
Eq.~\eqref{eq:a-cr} as follows:
\begin{equation} \label{eq:F4a-cr}
\tag{\ref{eq:a-cr}$^\prime$} %
F(a,\lambda) = \int_{\frac{a}{\lambda l_{\text{ex}}}}^\infty
\mathrm{d}x\ xe^{-2x^2}\left( 4x^2 - \frac{1}{x^2} +
\lambda^2\right)=0,
\end{equation}
which is an implicit form of the dependence $a(\lambda)$. The
critical value $a_c$ can be calculated from the condition ${\partial
F}/{\partial\lambda}=0$; hence $a_c
={l_\mathrm{ex}}\sqrt{x_0-x_0^2}$, where $x_0$ is a positive root of
the equation $2e^{-x}+\mathrm{Ei}(-x)=0$. This results in
$x_0\approx0.1$ and $a_c\approx 0.3{l_\mathrm{ex}}$.

\section{Magnetostatic energy calculations}
\label{sec:appendix-MS}

Let us write down an expression for the vortex energy with account
of the non--local magnetostatic contribution \eqref{eq:W-MS-z}:
\begin{equation} \label{eq:W4ring-with-magnetostatic}
\begin{split}
W_{\text{IMV}} = \frac{4\pi l_{\text{ex}}^2}{R^2}
\int\limits_{\frac{a}{\lambda l_{\text{ex}}}}^{\frac{R}{\lambda
l_{\text{ex}}}} x\mathrm{d}x \Biggl[ & \frac{1}{x^2} + \mu^2
e^{-2x^2}\left( \frac{4x^2}{1-\mu^2 e^{-2x^2}} -
\frac{1}{x^2}\right)\\
& + \lambda^2 \mu^2\mathcal{K}\left(x,\frac{a}{\lambda
l_{\text{ex}}},\frac{R}{\lambda l_{\text{ex}}},\frac{h}{\lambda
l_{\text{ex}}}\right) \Biggr].
\end{split}
\end{equation}
Here the kernel of integral
\begin{equation} \label{eq:K}
\mathcal{K}(x,\xi,\eta,\zeta) = \frac{e^{-x^2}}{\zeta}\int_\xi^\eta
x'e^{-x'^2}\int_0^\infty(1-e^{-\zeta
t})J_0(xt)J_0(x't)\:\mathrm{d}t\:\mathrm{d}x'.
\end{equation}
Note that the kernel of the local model \eqref{eq:W4ring} has an
exponential decay,
$\mathcal{K}(x,\bullet,\bullet,\bullet)=\exp(-2x^2)$.

Let us analyse the case of small thickness $(\zeta\to0)$:
\begin{equation}\label{eq:kernelZeta0}
\mathcal{K}\left(x,\xi,\eta,\zeta\to0\right)=e^{-x^2}\int_\xi^\eta
x'e^{-x'^2}\int_0^\infty tJ_0(xt)J_0(x't)\:\mathrm{d}t\:\mathrm{d}x'
\end{equation}
The internal integral in (\ref{eq:kernelZeta0}) can be calculated
directly:
\begin{equation}
\begin{split}\label{eq:Bessel_Delta}
&\int\limits_0^\infty tJ_0(at)J_0(bt)\mathrm
dt=\lim_{t\rightarrow\infty}\frac{t}{a^2-b^2}\left[a J_0(b t) J_1(a
t)-b
J_0(a t)J_1(b t)\right]=\\
=&\lim_{t\rightarrow\infty}\frac{1}{\sqrt{ab}}\left\{\cfrac{\sin(a-b)t}{\pi(a-b)}-\cfrac{\cos(a+b)t}{\pi(a+b)}\right\}
=\cfrac{1}{\sqrt{a b}}\left[\delta(a-b)+i\delta(a+b)\right]
\end{split}
\end{equation}
Finally the kernel (\ref{eq:kernelZeta0}) can be written in the form
\begin{equation}
\mathcal{K}\left(x,\xi,\eta,0\right)=e^{-x^2}\int_\xi^\eta
\sqrt{\cfrac{x'}{x}}\:e^{-x'^2}\delta(x'-x)\mathrm{d}x'=e^{-2x^2}.
\end{equation}
Thus the nonlocal model of magnetostatic interaction passes into
local model in the limit of thin films.

An account of nonlocal magnetostatic interaction changes the value
of the critical inner radius, when the magnetic phase transition
into the pure IPV phase occurs. This critical radius can be
calculated numerically as a solution of the following equation:
\begin{equation} \label{eq:a-cr-magnetostatic}
\int_{\frac{a}{\lambda l_{\text{ex}}}}^\infty \mathrm{d}x\ x\Biggl[
e^{-2x^2}\left( 4x^2 - \frac{1}{x^2}\right) +
\lambda^2\mathcal{K}\left(x,\frac{a}{\lambda
l_{\text{ex}}},\infty,\frac{h}{\lambda l_{\text{ex}}}\right)
\Biggr]=0
\end{equation}

To simplify the numerical calculations one can reduce the kernel to
\begin{equation}
\begin{split}
\label{kernelNew}
\mathcal{K}(x,\xi,\infty,\zeta)=&\cfrac{2\:e^{-x^2}}{\pi\zeta}
\Biggl\{\int_\xi^x\frac{t e^{-t^2}\mathrm K\left(t/x\right)}{x}
\mathrm dt+\int_x^\infty e^{-t^2}\mathrm K\left(x/t\right)\mathrm dt\\
&-\int_\xi^\infty \cfrac{te^{-t^2} \mathrm
K\left({2\sqrt{tx}}/{\sqrt{\zeta^2+(x+t)^2}}\right)}{\sqrt{\zeta^2+(x+t)^2}}
\:\mathrm dt\Biggr\},
\end{split}
\end{equation}
where $\mathrm{K}(x)$ is the elliptical integral of the first kind.

\begin{figure}
\begin{center}
\includegraphics[width=7.5cm]{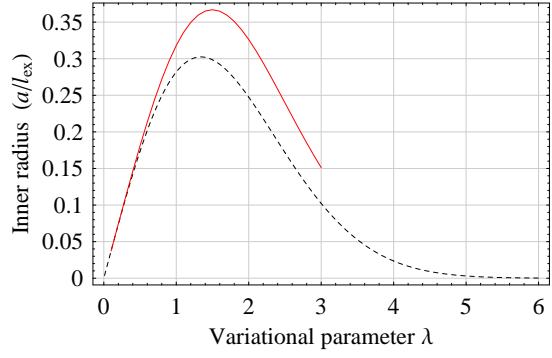}
\caption{The numerical solution of the
Eq.~\ref{eq:a-cr-magnetostatic} is denoted by solid red line. During
the calculation it was assumed that parameter
$\frac{h}{l_\mathrm{ex}}=\frac{5\:\mathrm{nm}}{5.3\:\mathrm{nm}}$,
that corresponds to the parameters of the simulations (see
Fig.~\ref{fig:vAmpl}). Dashed black line corresponds to the solution
of the model of local magnetostatics (\ref{eq:a-cr}),
Fig.~\ref{fig:acrVSlambda}.} \label{fig:acrVSlambdaNonlocal}
\end{center}
\end{figure}

The solution of the Eq.~(\ref{eq:a-cr-magnetostatic}) with kernel
(\ref{kernelNew}) is shown in Fig.~\ref{fig:acrVSlambdaNonlocal}.
The maximum of dependence $a(\lambda)$ that corresponds to the
critical value of inner radius is $a_c\approx 0.37 l_\mathrm{ex}$.
For the Py nanoring $a_c\approx 1.94\:\mathrm{nm}$, which agrees
with the simulations data, see Fig.~\ref{fig:vAmpl}.

\begin{figure}[h]
\begin{center}
\includegraphics[width=7.5cm]{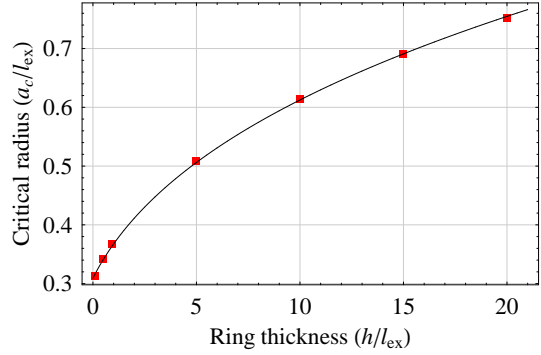}
\caption{The dependence of the critical inner radius on thickness of
the ring. Red bars denotes the exact values, which were obtained
numerically. Solid black line corresponds to the fit (see
text).}\label{A_cr_VS_h}
\end{center}
\end{figure}

\begin{figure}[h]
\begin{center}
\includegraphics[width=7.5cm]{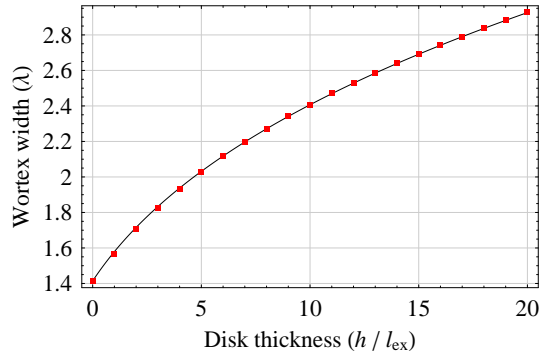}
\caption{The dependence of the vortex width in case of disk on disk
thickness.}\label{lambda_VS_h}
\end{center}
\end{figure}

An account of nonlocal interaction gives a possibility to calculate
the critical inner radius on thickness of the ring, see
Fig.~\ref{A_cr_VS_h}. This dependence can be fitted by the function
\begin{equation} \label{eq:a(h)-fit}
a_c^{\text{fit}}(h)\approx a_c\sqrt[3]{1 + {ch}/{l_{\text{ex}}}},
\qquad c \approx 0.68,
\end{equation}
where $a_c\approx 0.37 l_{\text{ex}}$ is the limit value for
infinitesimally thin rings, which was calculated in
Sec.~\ref{sec:vortex}. We also have numerically calculated the
dependence of vortex width ($\lambda$) on disk thickness (for
nonlocal magnetostatic model), see Fig.~\ref{lambda_VS_h}.

This dependence also can be well fitted by the third-root function:
\begin{equation} \label{eq:lambda(h)-fit}
\lambda^{\text{fit}}(h)\approx \sqrt{2}\:\sqrt[3]{1 +
{dh}/{l_{\text{ex}}}}, \qquad d \approx 0.39.
\end{equation}

{\scriptsize


}
\end{document}